\documentstyle[prb,aps,twocolumn,floats]{revtex}
\begin{document}
\twocolumn[\hsize\textwidth\columnwidth\hsize\csname@twocolumnfalse\endcsname
\draft
\title{Symmetry characterization of eigenstates in opal-based photonic
crystals} \author{F.\ L\'opez-Tejeira$^\dag$, T. Ochiai} 
\address{Departamento de F\'{\i}sica Te\'orica de la Materia Condensada,
 Facultad de Ciencias,\\ Universidad Aut\'onoma de Madrid,
28049 Madrid, Spain.}
\author{K.\ Sakoda}
\address{Research Institute for Electronic Science, Hokkaido University,
  \ North \ 12 West 6, Kita-ku,\\ Sapporo 060-0812, Japan}
\author{J.\ S\'anchez-Dehesa}
\address{Departamento de F\'{\i}sica Te\'orica de la Materia Condensada,
 Facultad de Ciencias,\\ Universidad Aut\'onoma de Madrid,
28049 Madrid, Spain.}

\date{\today}

\maketitle
\begin{abstract}
The complete symmetry characterization of eigenstates in bare opal systems is obtained by means of
group theory. This symmetry assignment has allowed us to identify several bands
that cannot couple with an incident external plane wave. Our prediction is supported by 
layer-KKR calculations, which are also performed: the coupling coefficients between bulk
modes and externally excited field tend to zero when symmetry properties mismatch.
\end{abstract}
]

\section{Introduction}

Periodic dielectric structures (the so called photonic crystals) are 
probably one of the most exciting topics in contemporary physics, not only for its 
scientific relevance but also because of  the remarkable applications that have 
been suggested, specially for  systems with gaps in the visible and infrared 
range. So, since the first proposal in the late eighties,\cite{yablo,john} many
efforts have been devoted to the fabrication of  materials with an 
absolute frequency gap in such region. However, material engineering of 3D 
periodic systems at the relevant length scale (hundreds of nm to 1 micron) 
presents non trivial technological problems in a straightforward manufacture.
Several different self-assembly approaches have been proposed\cite{selfa} and
one of the most promising strategies is that based on artificial opal
systems.\cite{adv97,vlas}

It is clear that one of the most important aspects of sample characterization relies on a
proper understanding of optical measurements. The structures within the reflectance or
transmittance spectra are currently related with gaps and pseudogaps in the infinite band
structure, but the symmetry properties of its eigenfunctions can also produce unexpected
patterns: symmetry-inactive modes (ie. bands that cannot be coupled with incident light
due to symmetry reasons) have been reported of some 2D photonic crystals
\cite{rober,sako,krauss} and some authors have predicted the same phenomenon to take place
in 3D materials as well.\cite{stefa,ohta,sakogr,kara,yuan}
Thus, the classification of eigenmodes according to their symmetry properties can provide
us valuable information when comparing our calculation with experimental results.

The aim of this work is to present a complete symmetry characterization of eigenstates for
the abovementioned opal structure. In order to reach this goal, the paper is organized as follows: 
In Section II we introduce the solution of Maxwell equations as an eigenvalue problem that can be
restricted to a band structure calculation. Section III describes the symmetry properties of
electric field according to group theory. In Section IV we present the mode assignment for a
bare opal system. The existence of uncoupled modes is discussed in Section V. Finally, in Section VI
we summarize our work.

\section{Maxwell equations and photonic band structure}

As far as the point at stake is the propagation of light in a dielectric
material, we shall concern ourselves with the macroscopic Maxwell equations. If
we impose our problem to fulfil some requirements,\cite{bookjoann} we can
decouple the equations in order to finally arrive to an expression entirely
in terms of the magnetic field \(\mathbf{H}(\mathbf{r})\)

\begin{equation}
\label{eqmax}
\nabla  \times \left( \frac{1}{\varepsilon (\mathbf{r})}\nabla
\times\mathbf{H}(\mathbf{r})\right)=\left(
\frac{\omega}{c}\right)
^{2}\mathbf{H}(\mathbf{r})
\end{equation}
 
\noindent
With the definition of a suitable differential operator

\begin{equation}
\widehat{\mathrm{\Theta} }\mathbf{H}(\mathbf{r}) \equiv \nabla  \times
\left( \frac{\mathrm{1}}{\varepsilon (\mathbf{r})}\nabla
\times\mathbf{H}(\mathbf{r}) \right)  \end{equation}

\noindent
equation (\ref{eqmax}) can be explicitly written like an ordinary eigenvalue
problem:

\begin{equation}
\widehat{\Theta }\mathbf{H}(\mathbf{r})=\left(
\frac{\omega}{c}\right)
^{2}\mathbf{H}(\mathbf{r})
\label{opeH}
\end{equation}

\noindent
The somehow arbitray definition of \( \widehat{\Theta } \) ensures the
operator be hermitian, which implies that its
eigenfunctions have several properties that are
extremely useful in numerical calculations.\cite{bookjoann}

In the alternate approach, we can eliminate \(\mathbf{H}(\mathbf{r})\) to obtain
an equivalent formulation for the stationally electric field. This way of
considering the problem results in a generalized eigenvalue equation with the
dielectric function playing the role of a new hermitian
operator in the right side:

\begin{eqnarray}
\label{eqmax2}
\widehat{\Xi}_{1}\mathbf{E}(\mathbf{r})\equiv
\nabla  \times \left( \nabla \times\mathbf{E}(\mathbf{r})\right)=
\nonumber\\
=\left(\frac{\omega}{c}\right)^{2} \varepsilon
(\mathbf{r})\mathbf{E}(\mathbf{r})\equiv
\left(\frac{\omega}{c}\right)^{2}\mathrm{\widehat{\Xi}}_{2}\mathbf{E}(\mathbf{r})
\end{eqnarray}

It is clear that both choices provide the correct physics,
so the preference for (\ref{opeH}) instead of (\ref{eqmax2})
(or viceversa) in
the numerical evaluations will only depend on our personal convenience.

We finally want to remark that Bloch's theorem can be applied to photonic
crystals due to the periodicity of the dielectric function
\(\varepsilon(\textbf{r})\).
Thus, we can restrict the eigenvalue problem
to the calculation of the band structure of our system and therefore
rewrite Equations (\ref{opeH}) and (\ref{eqmax2}) as

\begin{equation}
\label{eqBloch1}
\widehat{\Theta }\mathbf{H}^{\mathit{n}}_{\mathbf{k}}(\mathbf{r})=\left(
\frac{\omega_{\small{\mathit{n}}}(\small{\mathbf{k}})  }{\mathrm{c}}\right)
^{2}\mathbf{H}^{\mathit{n}}_{\mathbf{k}}(\mathbf{r})
\end{equation}

\begin{equation}
\label{eqBloch2}
\widehat{\Xi}_{1}\mathbf{E}^{\mathit{n}}_{\mathbf{k}}(\mathbf{r})=\left(
\frac{\omega_{\small{\mathit{n}}}(\small{\mathbf{k}})  }{\mathrm{c}}\right)
^{2}\mathrm{\widehat{\Xi}}_{2}\mathbf{E}^{\mathit{n}}_{\mathbf{k}}(\mathbf{r})
\end{equation}

\noindent
where \( \omega  \) is the frequency of  the eigenstates, \(\mathbf{k}\) the Bloch
wave vector within the first Brillouin zone and \(\mathit{n}\) a discrete index
for increasing eigenfrequencies. 

\section{Eigenmode characterization}
\subsection{Eigenfunctions and symmetry operators}

Bloch's theorem states that the electric field within a photonic crystal
can be written as 

\begin{equation}
\label{eqBlochE}
\mathbf{E}_{\mathbf{k}}^{\mathit{n}}(\mathbf{r})=e^{\mathit{i}\mathbf{k \cdot
r}}\mathbf{u}_{\mathbf{k}}(\mathbf{r})
\end{equation}
\noindent
in which  \(\mathbf{k}\) is the Bloch wave vector inside the first
Brillouin zone and \(\mathbf{u}_{\mathbf{k}}(\mathbf{r})\)  a periodic
vector function of the lattice structure.
It follows from the abovementioned periodicity that

\begin{equation}
\label{eqBlochE1}
\mathbf{E}_{\mathbf{k}}^{\mathit{n}}=e^{\mathit{i}\mathbf{k \cdot r }}
\sum_{\mathbf{q}}\mathbf{f}_{\mathbf{q}}e^{\mathit{i}\mathbf{q \cdot r}} 
\end{equation}

\noindent
where \(\mathbf{q}\) is a reciprocal lattice vector of the structure.

Let \( \widehat{A} \) be a symmetry operator of the lattice. The way in
which a general symmetry transformation \( \widehat{A} \)
operates over a vector field \(\mathbf{f}(\mathbf{r})\) is:

\begin{equation}
\label{eqsim}
\widehat{A}\bullet \mathbf{f}(\mathbf{r})=Af(A^{-1}\mathbf{r})
\end{equation}    

\pagebreak
\noindent
Hence

\begin{equation}
\label{eqBlochE2}
\widehat{A} \bullet
\mathbf{E}_{\mathbf{k}}^{\mathit{n}}=Ae^{\mathit{i}\mathbf{k \cdot
A^{-1}r }}
\sum_{\mathbf{q}}\mathbf{f_{q}}e^{\mathit{i}\mathbf{q \cdot A^{-1}r }}
\end{equation}         
   
\noindent
Taking the orthogonality of symmetry matrices into account,

\begin{equation}
\label{eqBlochE3}
\widehat{A} \bullet
\mathbf{E}_{\mathbf{k}}^{\mathit{n}}=Ae^{\mathit{i}\mathbf{Ak \cdot
r }}
\sum_{\mathbf{q}}\mathbf{f_{q}}e^{\mathit{i}\mathbf{Aq \cdot r }}
\end{equation}
For  a wave vector \(\mathbf{k}\), the \(\mathbf{k}\)-vector 
point group \(G_{\mathbf{k}}\) is defined as the set of symmetry operations
which satisfy

\begin{equation}
\label{kgroup}
\mathbf{Ak}=\mathbf{k}+\mathbf{q'}
\end {equation}
\noindent
with \(\mathbf{q'}\) in the reciprocal lattice. Thus, \(\widehat{A}\)
being part of \(G_{\mathbf{k}}\),

\begin{equation}
\label{eqBlochE4}
\widehat{A} \bullet
\mathbf{E}_{\mathbf{k}}^{\mathit{n}}=e^{\mathit{i}\mathbf{(k+q') \cdot
r }}
\sum_{\mathbf{q}}\mathbf{Af_{q}}e^{\mathit{i}\mathbf{Aq \cdot r }}
\end{equation} 

\noindent
Given that the extra phase can always be included in the infinite
summatory, we conclude that the symmetry transformation of
\(\mathbf{E}_{\mathbf{k}}^{\mathit{n}}(\mathbf{r})\) generates another
Bloch function with the same wave vector:

\begin{equation}
\label{eqBlochE5}
\widehat{A} \bullet
\mathbf{E}_{\mathbf{k}}^{\mathit{n}}=e^{\mathit{i}\mathbf{k \cdot
r }}\mathbf{v_{k}}(\mathbf{r})
\end{equation}
Let us now consider the action of a symmetry operator \( \widehat{A} \) on
both sides of Equation (\ref{eqBloch2}):

\begin{equation}
\label{AeqBloch2}
(\widehat{A}\bullet
\widehat{\Xi}_{1})\mathbf{E}^{\mathit{n}}_{\mathbf{k}}(\mathbf{r})=\left(
\frac{\omega_{\small{\mathit{n}}}(\small{\mathbf{k}})  }{c}\right)
^{2}(\mathrm{\widehat{A} \bullet \widehat{\Xi}_{2}})
\mathbf{E}^{\mathit{n}}_{\mathbf{k}}(\mathbf{r})
\end{equation}

\noindent
For any symmetry operator that leaves \(\varepsilon(\textbf{r})  \) invariant
within the unit cell, it can be easily proved  that it commutes with \(
\widehat{\Xi}_{1} \) and \(\widehat{\Xi}_{2} \). Therefore 

\begin{equation}
\label{A2eqBloch2}
\widehat{\Xi}_{1}(\widehat{A}
\bullet \mathbf{E}^{\mathit{n}}_{\mathbf{k}}(\mathbf{r}))=\left(
\frac{\omega_{\small{\mathit{n}}}(\small{\mathbf{k}})  }{c}\right)
^{2} \mathrm{\widehat{\Xi}_{2}} (\widehat{A} \bullet 
\mathbf{E}^{\mathit{n}}_{\mathbf{k}}(\mathbf{r})) \end{equation}

\noindent
which implies both \(\mathbf{E}_{\mathbf{k}}^{\mathit{n}}(\mathbf{r}) \) and 
\((\widehat{A} \bullet \mathbf{E}_{\mathbf{k}}^{\mathit{n}}(\mathbf{r})) \) be
eigenfunctions with the same eigenfrequency.

On the assumption that \( G_{\mathbf{k}}=\{\widehat{A}_{i}\} \) is the
point group of some wave vector \textbf{k}, it follows from
Equations (\ref{eqBlochE5}) and (\ref{A2eqBloch2}) that
\(\mathbf{E}_{\mathbf{k}}^{\mathit{n}}(\mathbf{r}) \) must verify
 
\begin{equation}
\label{Eqvieja}
\widehat{A}_{i}\mathbf{E}_{\mathbf{k},\mathrm{g}}^{\mathit{n}}(\mathbf{r})=
\sum _{\mathrm{g'}}\alpha ^{\mathrm{i}}_{\mathrm{g,
g'}}\mathbf{E}_{\mathbf{k},\mathrm{g'}}^{\mathit{n}}(\mathbf{r})\\
\linebreak
\end{equation}

\noindent
where \(g\) is eigenmode degeneracy. We shall henceforth concern
ourselves with the meaning of that set of scalar
coefficients \( \{\alpha ^{i}_{\mathrm{g, g'}}\} \).

\subsection{Theory of representations} 

Within the framework of the usual group theory,\cite{fal} it can be straightforwardly
proved that the scalar coefficient matrices \( \widetilde{\alpha}^i \) in
(\ref{Eqvieja}) form a representation of (i.e. are homomorphic with) \(
G_{\mathbf{k}}\) group. On the other hand, in the case where
the degeneracy is said to be normal (i.e. when it is not possible to remove
such a degeneracy by changing a parameter which does not alter the stated
symmetry of the problem), it may be seen that the set \(\{\mathbf{E}_{\mathbf{k},
\mathrm{g}}^{\mathit{n}}(\mathbf{r})\}\) spans an irreducible invariant space
under the \(G_{\mathbf{k}}\) group of symmetry operators. Thus, the \(
\{\widetilde{\alpha}^i\} \) set of matrixes constitutes an irreducible
representation \( R\) of \(G_{\mathbf{k}}\) point group, and \( R \) label 
can be therefore assigned to \textit{n}th eigenmode at \textbf{k} point.

If \( \widehat{A}_{i} \) 
belongs
to a class \( C \), then

\begin{equation}
\label{eqtraza}
\sum _{g}\alpha ^{i}_{g, g}=\chi ^{\, \! R}_{C}
\end{equation}

\noindent
with \( \chi ^{\, \! R}_{C} \) the character of class \( C \) in some 
irreducible representation \( R \). Given that each \( R \) has its own
characters, we can use them to distinguish the irreducible representations in
order to classify the different eigenmodes.\cite{booksako} Further details about character
evaluation can be found in Appendix A.

\section{Symmetry Characterization in Bare Opal Systems}

Synthetic bare opals are constituted by SiO\( _{2} \) spheres that
organize themselves to form a face-centered cubic (fcc) lattice. As far as this
process has been extensively described in previous works\cite{adv97}, we just
mention that opaline sample growth method is based on natural sedimentation of
silica nanospheres along the (111) direction of fcc lattice. Although
the refractive index contrast between  SiO\( _{2} \) and air does not allow
bare opals to exhibit complete gaps, they can be  used as templates to
obtain the so called inverted structures,\cite{vos,zak,lang00,blanco,resina}
which are one of the most promising strategies between the
different self-assembly approaches to photonic crystal fabrication.
Thus, no one can deny that bare opals still constitute a topic of present interest.

In order to work out the classification of eigenstates described in the previous section, we
shall first concern ourselves with the intrinsic properties of the face-centered
cubic lattice, as far as the symmetry point group which determines the set of allowed
irreducible representations for any wave vector \textbf{k} is irrespective
of the particular realization of our system. Table I lists the symmetry point groups
at the corners of the irreducible part of the first Brillouin zone (see Figure 1) according to
Schoenflies notation. Given that \textit{R} representation can be assigned to \textit{n}th
mode at one of those particular points, we can then easily derive all its compatibility
relations (see Appendix B) with the corresponding modes at adjacent wave vectors by
following the loss of symmetry as we go from one point group to another which is a
subgroup of the first.

In Figures 2 to 4 we present the photonic band structure of a face-centered cubic lattice of
spheres. Results are plotted in terms of reduced frequency \( a/\lambda \) where \(a\)
and \(\lambda\) denote the lattice parameter and the vacuum wavelength respectively.         
The dielectric constants of the sphere and the background were assumed to be 2.104
(silica) and 1.0 (air). The ratio of the lattice parameter and the sphere radius  was
chosen to verify close-packing condition. This band calculation was carried out by means of
an iterative implementation\cite{mpb} of the plane wave expansion method on Equation (\ref{opeH}). 
Of course, the photonic band structure of a bare opal system has been previously
calculated,\cite{reyn} so the main outcome of our present work is obviously the
mode assignment for the electric field. The labels within the figures have been 
determined by the  numerical evaluation of only several eigenmodes, as far as the abovementioned
compatibility relations allowed us to easily connect the irreducible representations for
adjacent wave vectors. For the sake of completeness, we also list in Table II the symmetry
characterization of eigenmodes at \(\Gamma\), \( X\), \(L\), \(U\), \(W \) and \(K\) points
for increasing values of the eigenfrequency.

\section{Uncoupled Bands}
\subsection{Symmetry assignment and uncoupled modes}
As mentioned in Sec.I, light attenuation produced  by uncoupled bands 
( i.e., bands with eigenmodes that cannot be excited by the incident light because of symmetry
reasons) has been reported of some 2D photonic crystals. Moreover, some authors have predicted a
similar phenomenon to also take place in 3D materials,\cite{stefa,ohta,sakogr,kara,yuan} but as far
as we know, there is no experimental evidence of this  mechanism operating in an
opaline structure.
The exposition of how these  uncoupled modes can be identified as a simple consecuence of the
symmetry assignment will be the aim of the present section.

In order to reach this goal, let us focus our interest on what happens when light incidence
is normal to the \{111\} set of planes in the opal. This is the direction along the
structure naturally grows and from which experimental data are usually obtained.
If we consider the \(z\) axis in the normal direction  to the
sample surface, the incident plane waves can be expressed in terms of 
the following vector basis

\begin{equation}
\{\mathbf v_{1} \mathrm = (1,0,0)\mathit e^{ikz} ;
\hspace{1mm} \mathbf v_{2} \mathrm = (0,1,0) \mathit e^{ikz}\} 
\label{eqbasis}
\end{equation}

\noindent
where \(k\) is the wave number of the plane waves in free space.
On the other hand, it is clear that vectors in (\ref{eqbasis})
also form a  basis set of some reducible or irreducible representation 
of \(C_{3v}\) point group. In order to determine that representation, we 
have to apply the symmetry operations of  \(C_{3v}\) over the vector basis:

\begin{equation}
\left\{ \begin{array}{c}
\widehat{C}_{3z} \mathbf v_{1}\mathrm=-\frac{1}{2}
\mathbf v_{1} \mathrm +\frac{\sqrt{3}}{2}\mathbf v_{2}\\
\widehat{C}_{3z}\mathbf v_{2} \mathrm
=-\frac{\sqrt{3}}{2} \mathbf v_{1} \mathrm -\frac{1}{2}\mathbf v_{2}
\end{array}\right.
\label{equc3}
\end{equation}
\begin{equation}
\left\{ \begin{array}{l}
\widehat{\sigma }_{v} \mathbf v_{1} \mathrm = - \mathbf v_{1} \\
\widehat{\sigma }_{v} \mathbf v_{2} \mathrm =\hspace{2.6mm}\mathbf 
v_{2}
\end{array}\right.
\label{equsig}
\end{equation}

From Eqs. (\ref{equc3}) and (\ref{equsig}) we can obtain the trace (character) 
of the transformation matrixes:

\begin{equation}
\label{equtr}
\mathit \chi (C_{3z})=\mathrm -1;\hspace{2mm}\mathit 
\chi(\sigma_{v})= \mathrm 0 
\end{equation}

\noindent
If we compare these results with the properties of \(C_{3v}\) point 
group, we find that these are the characters of the \(E\) representation. 
But the boundary condition of tangential components of EM field being
continous at the interface imposes the representation (i.e. the symmetry
properties) both inside and outside the system to be the same. 
Consequently, only \( E\)-labelled states will contribute to light transport 
along the (111) direction of the structure. This kind of restriction in available 
bands is present in several  directions of the lattice, but we are mainly interested on
those which include the \(\Gamma\) point, i.e. those of which some observable
consecuences are expected.

In Table III we summarize the information about uncoupled modes along the relevant
directions of a face-centered cubic structure provided by symmetry considerations.
Let there be a frequency range where only uncoupled modes exist. Despite of
its nonzero density of states, we would expect a total reflection (zero
transmission) in such a range as a consequence of photonic states not  being
coupled with the incidence plane wave. 

Unfortunately, this is not the case for bare opal systems:
as shown in Figures 2 to 4, uncoupled modes always coexist with available
ones, and consequently transmittance measurements cannot confirm the predictions of
group theory. Therefore, an alternative test will be required.

\subsection{Estimation of the coupling by means of the layer-KKR method}
It is well known that the  photonic band structure of a regular array of non-overlapping spheres 
in a host medium can be very accurately calculated within the framework of the layer-KKR
method\cite{stefa,stefa2,ohta2,ohta3,ohta4,modi}. Moreover, we can also obtain
the transmission and reflection coefficients through a finite slab of such material 
combining the scattering matrices of its different layers.
On the assumption that slab thickness is very large, the electromagnetic field in the 
middle of that slab can be regarded as a superposition of the true bulk eigenstates.
Thus, a kind of overlapping integral of this field and the actual eigenstates
of the infinite system would provide an estimation of the coupling between the structure and
the incident light. 

Let the $z$ axis be in the normal direction to the slab surface. Hence, the electric field
between two consecutive sphere layers can be expressed in terms of its Fourier components as

\begin{eqnarray}
{\bf E}_{\rm void}({\bf r})=
\sum_{\bf q}{\bf u}_q^{+}e^{i{\bf K}_q^+\cdot{\bf r}}
    + {\bf u}_{q}^-e^{i{\bf K}_q^-\cdot{\bf r}}& \\
{\bf K}_q^{\pm}={\bf k}_{\|}+{\bf q}\pm \hat{z}
\sqrt{{\omega^2\over c^2}-({\bf k}_{\|}+{\bf q})^2}&
\end{eqnarray}
\noindent
Here,  ${\bf q}$ denotes the 2D reciprocal lattice vector associated
to the surface periodicity, and $k_{\|} \equiv (k_x, k_y)$  the
wave vector within the  surface Brillouin zone.
On the other hand, these Fourier components also verify

\begin{equation}
\mathit{Q}\left(\begin{array}{c}
{\bf u}_q^+\\
{\bf u}_q^-
\end{array}
\right) = e^{ik_zz}\left(\begin{array}{c}
{\bf u}_q^+\\
{\bf u}_q^-
\end{array}
\right)
\end{equation}
\noindent
where $Q$ is an appropriately constructed tranfer matrix which relates
the electric field in the void between the $(N-1)$-th and 
$N$-th layers to that between  the $N$-th and $(N+1)$-th ones.
Since $Q$ is not hermitian, its left eigenstate 

\begin{equation}
u_L=\left(\begin{array}{c}
{\bf u}_q^+\\
{\bf u}_q^-
\end{array}
\right)
\end{equation}
\noindent
is generally different from the complex conjugate of the right one,
but we assume that both are normalized  to meet the orthogonality condition:
\begin{equation}
 (u_L^n)^\dagger u_R^{n'}   = \delta_{nn'}.
\end{equation}
We can then define the coupling coefficient $c_n$ as
\begin{equation}
c_n= (u_L^n)^\dagger u_{\rm bulk},
\label{coe}
\end{equation}
with $u_{\rm bulk}$ a column vector which contains the Fourier coefficients of 
the electric field excited by the incident light in the bulk part of the
system.\cite{tetsu1}
Notice that  Equation (\ref{coe}) involves a selection rule on the spatial symmetry
of the electric field.
Since the transfer matrix $Q$ commutes with the symmetry operations of the
point group relevant to the incidence surface, the eigenstates of $Q$ are classified
according to the irreducible representations of that point group.
Hence, $c_n$ must become zero if  $u_{\rm bulk}$ is attributed
to a different irreducible representation from that of  $u_L^n$.

In order to confront our estimation with the predictions of group theory, let us
consider again the previous example of normal incidence along the (111) direction. 
In this case incident light has a \(k_{\parallel}= 0\) component within the surface
Brillouin zone associated with the \{111\} set of planes. Following the abovementioned
procedure, we employed a 32-monolayer bare opal slab to obtain the electric field distribution in the space between the 16th and 17th layers, where the bulk configuration is assumed
to be achieved. At the same time, the photonic dispersion relation of the infinite periodic
system was also re-calculated for increasing values of \(k_{z}\).
Figure 5 shows both dispersion relation and symmetry assignment within the [1.10,1.3]
interval. Now, we  will pay particular atention to the pair of frequency values marked
with horizontal lines: for one of them ($a/\lambda=1.15$) no uncoupled bands are predicted
by group theory, while for the other ($a/\lambda=1.25$) some of that bands are expected.
In the first case, we obtained $|c_{n}|^2$ equal to 0.689 and 0.348
for $E$ bands with positive group velocity and $|c_{n}|^2=0.007, 0.006$
for the ones with negative slope. When calculated for the higher frecuency,
the values of $|c_{n}|^2$  for $E(+)$ and $E(-)$ bands  are
\{0.265, 0.377, 0.109, 0.035\} and
\{0.085, 0.016, 0.098, 0.054\} respectively.
With regards to the uncoupled bands (i.e, those with $A_1$ symmetry),
their coefficients are less than $10^{-16}$. Therefore we can conclude that light
transmission is actually forbidden for such modes, as far as they do not contribute to the
excited field inside the slab. This constitutes a confirmation of group theory predictions.

\section{Summary}

We have analyzed the symmetry properties of eigenstates along the high symmetry 
directions of close-packed bare opals according to the group theory.
We found that some bands cannot be coupled with an external plane
wave along several directions because of symmetry reasons, which was
confirmed by layer-KKR calculations; nevertheless, the presence of uncoupled
modes is not expected to provide any observable features in the  transmittance
for this particular system.

\section*{Acknowledgments}

The support of the European Commission, project IST-1999-19009 PHOBOS,
is gratefully acknowledged.
 We also acknowledge financial aid provided by the Spanish CICYT 
 (Project No. MAT2000-1670-C04-04).
F.L.T. and J.S.D. acknowledge  
 the computing facilities provided by the {\it Centro de Computaci\'on 
Cient\'{\i}fica} at the UAM.

\appendix
\section{Numerical assignment of irreducible representations}

As stated in Section III, the label assignment for a given mode
\(\mathbf{E}^{\mathit{n}}_{\mathbf{k}}(\mathbf{r})\) is based on the evaluation of its
character under the symmetry operators within \(G_{\mathbf{k}}\). The first thing that
needs to be remarked is that solutions obtained by means of the plane wave expansion
method (in fact by means of any numerical calculation) do not satisfy Equation
(\ref{Eqvieja}) but in an approximate way. However, a suitable algorithm can always be
found in order to estimate the character of computed eigenfunctions.

In our present work, we have defined a strategy based on direct evaluation of Equation
(\ref{Eqvieja}), which will be now rewritten  for the sake of clarity:

\begin{equation}
\label{Eqvieja2}
\mathbf{P}_{\mathrm{g}}\equiv\widehat{A}\mathbf{E}_{\mathrm{g}}=
\sum _{\mathrm{g'}}\alpha_{\mathrm{g,g'}}\mathbf{E}_{\mathrm{g'}}
\end{equation}

\noindent
In the case of one-dimensional irreducible representations (\(g=1\)), the character
\(\chi(\widehat{A})\) is given by

\begin{equation}
\label{char1D}
\chi_{1D}=\alpha=\frac{\mathbf{P}\cdot\mathbf{E}}{|\mathbf{E}|^2}
\end{equation}

\noindent
Similar expressions can  be obtained for two-dimensional (\(g \le 2\))
representations

\begin{equation}
\label{char2D}
\chi_{2D}=\sum_{j,k}
\frac{(-1)^{j+k}[(\mathbf{P_{j}}\cdot\mathbf{E_{k}})(\mathbf{E_{k}}\cdot\mathbf{E_{j}})-(\mathbf{P_{j}}\cdot\mathbf{E_{j}})|\mathbf{E_{k}}|^2]}{|\mathbf{E_{1}}|^2|\mathbf{E_{2}}|^2-(\mathbf{E_{1}}\cdot\mathbf{E_{2}})(\mathbf{E_{2}}\cdot\mathbf{E_{1}})}
\end{equation}

\noindent
and also for three-dimensional (\(g \le 3\)) ones, given that  some auxiliary
quantities be  defined:

\begin{equation}
\label{char3D}
\chi_{3D}=\frac{\mathbf{w_{1}\cdot(v_{2}\times v_{3})+
v_{1}\cdot(w_{2}\times v_{3})+v_{1}\cdot(v_{2}\times w_{3})}}
{\mathbf{v_{1}\cdot(v_{2}\times v_{3})}}
\end{equation}

\begin{eqnarray}
(\mathbf{v_{k}})_{j}&=&\mathbf{E_{k}\cdot E_{j}}\\
(\mathbf{w_{k}})_{j}&=&\mathbf{P_{k}\cdot E_{j}}
\end{eqnarray}

We finally have to mention that all symmetry assignments have been confirmed
by means of a different numerical method for character evaluation
which was  proposed in a previous paper.\cite{tetsu2}

\section{Compatibility relations of irreducible representations in the fcc lattice}

In Tables IV to IX, the compatibility relations between the irreducible
representations for the points at the corners of the irreducible part of the fcc 
first Brillouin zone and those for the wave vectors on intermediate segments are presented.
It can be observed that only one mirror reflection is defined on many of these segments
because of their low symmetry invariance.

\bibliographystyle{prsty}

\begin{thebibliography}{}
\bibitem[\dag]{byline}Author to whom correspondence should
be addressed: fernando.lopeztejeira@uam.es
\bibitem{yablo}E. Yablonovitch, \emph{Phys. Rev. Lett.} \textbf{58}, 2059 (1987)
\bibitem{john}S. John, \emph{Phys. Rev. Lett.} \textbf{58}, 2486 (1987)
\bibitem{selfa}See for instance L. Mart\'{\i}n-Moreno, F.J. Garc\'{\i}a-Vidal
and
A.M. Somoza, \emph{Phys. Rev. Lett.} \textbf{83}, 73 (1999) and references
therein. A more recent review can be found in Y. Xia, B. Gates and
Z.Y. Li, \emph{Adv. Mater.} \textbf{13}, 409 (2001)
\bibitem{adv97}R. Mayoral, J. Requena, S.J. Moya, C. L\'{o}pez, A. Cintas, H.
M\'{\i}guez,
F. Meseguer, L. V\'{a}zquez, M. Holgado and A. Blanco, \emph{Adv.
Mater.} \textbf{9}, 257 (1997)
\bibitem{vlas}Yu. A. Vlasov, V.N. Astratov, O.Z. Karimov, A.A. Kaplyanskii, V.N.
Bogomolov and A.V. Prokofiev, \emph{Phys. Rev. B} \textbf{55}, 13357
(1997)
\bibitem{rober}W. M. Robertson, G. Arjavalingam, R.D. Meade, K.D. Brommer, A.M. Rappe
and J.D. Joannopoulos, \emph{Phys. Rev. Lett.} \textbf{68}, 2023 (1992)
\bibitem{sako}K. Sakoda, \emph{Phys. Rev. B} \textbf{52}, 7982 (1995)
\bibitem{krauss}T.F. Krauss, R.M. de la Rue, and S. Brand, \emph{Nature} \textbf{383},
699 (1996)
\bibitem{stefa}N. Stefanou, V. Karathanos and A. Modinos, \emph{J. Phys: Condens.
Matter} \textbf{4}, 7389 (1992)
\bibitem{ohta}K. Ohtaka and Y. Tanabe, \emph{J. Phys. Soc. Jpn.} \textbf{65}, 2670
(1996)
\bibitem{sakogr}K. Sakoda, \emph{Phys. Rev B} \textbf{55}, 15345 (1997)
\bibitem{kara}V. Karathanos, \emph{J. Mod. Opt.} \textbf{48}, 1751 (1998)
\bibitem{yuan}Z. Yuan, J.W. Haus and K. Sakoda, \emph{Opt. Exp.} \textbf{3}, 19
(1998)
\bibitem{bookjoann}J.D. Joannopoulos, R.D. Meade and J.W. Win, \emph{Photonic Crystals},
(Princeton University Press, Princeton 1995)
\bibitem{fal}See for instance L.M. Falicov, \emph{Group Theory and its Physical
Applications} ( The University of Chicago Press, Chicago, 1966)
\bibitem {booksako} A comprehensive investigation on the role of symmetry in photonic
crystals can be found in K. Sakoda, \emph{Optical Properties of Photonic Crystals},
(Springer Verlag, Berlin, 2001)
\bibitem{vos}J.E.G.J. Wijnhoven and W.L. Vos, \emph{Science} \textbf{281}, 802
(1998)
\bibitem{zak}A.A. Zakhidov, R.H. Baughman, Z. Iqbal, C. Cui, I. Khayrullin, S.O.
Dantas, J. Marti and V.G. Ralchendo, \emph{Science} \textbf{282},
897 (1998)
\bibitem{lang00}H. M\'{\i}guez, F. Meseguer, C. L\'{o}pez, M. Holgado, G. Andreasen,
A. Mifsud and V. Forn\'{e}s, \emph{Langmuir} \textbf{16}, 4405 (2000)
\bibitem{blanco}A. Blanco, E. Chomsky, S. Grabtchak, M. Ibisate, S. John, S.W. Leonard,
C. L\'{o}pez, F. Meseguer, H. M\'{\i}guez, J.P. Mondia, G. Ozin,
O. Toader and H.M. Van Driel, \emph{Nature} \textbf{405}, 437 (2000)
\bibitem{resina}H. M\'{\i}guez, F. Meseguer, C. L\'{o}pez, F. L\'{o}pez-Tejeira
and J. S\'{a}nchez-Dehesa, \emph{Adv. Mater.} \textbf{13}, 393 (2001)
\bibitem{mpb}S.G. Johnson and J.D. Joannopoulos, \emph{Opt. Exp.} \textbf{8},
173 (2001).
\bibitem{reyn}A. Reynolds, F. L\'{o}pez-Tejeira, D. Cassagne, F.J. Garc\'{\i}a-Vidal,
C. Jouanin and J. S\'{a}nchez-Dehesa, \emph{Phys. Rev. B} \textbf{60},
11422 (1999)
\bibitem{stefa2}N. Stefanou, V. Karathanos and A. Modinos, \emph{Comp. Phys. Commun.}
\textbf{113}, 49 (1998); ibid \textbf{132}, 189 (2000)
\bibitem{ohta2}K. Ohtaka,\emph{Phys. Rev. B}, \textbf{19}, 5057
(1979); \emph{J. Phys. C}, \textbf{13}, 667 (1980) 
\bibitem{ohta3}K. Ohtaka and Y. Tanabe, \emph{J. Phys. Soc. Jpn.} \textbf{65}, 2276 (1996)
\bibitem{ohta4}K. Ohtaka, T. Ueta and K. Amemiya, \emph{Phys. Rev. B}, \textbf{57}, 2550
(1998)
\pagebreak
\bibitem{modi}A. Modinos, N. Stefanou and V. Yannopapas, \emph{Opt. Exp.}, \textbf{8}, 197
(2001)
\bibitem{tetsu1}Further details can be found in T. Ochiai and J. S\'{a}nchez-Dehesa,
\emph{Phys. Rev. B} (in press)
\bibitem{tetsu2}T. Ochiai (submitted to \emph{Phys. Rev. B}) 
\end{thebibliography}

\textbf{Figure 1}.The first Brillouin zone of the fcc lattice. The lines between white
circles define the boundaries of the irreducible part of the zone.

\textbf{Figure 2}. Photonic band structure along the \(\overline{XU}\) and
\(\overline{UL}\) directions for a bare opal system. Labels refer to irreducible
representations of \(C_{2v}\) and \(C_{1h}\) respectively. An inset figure is included for
the sake of clarity.

\textbf{Figure 3}. Photonic band structure along the \(\overline{L \Gamma}\) and
\(\overline{\Gamma X}\) directions for a bare opal system. Labels refer to irreducible
representations of \(C_{3v}\) and \(C_{4v}\) respectively. Some inset figures are included
for the sake of clarity.

\textbf{Figure 4}. Photonic band structure along the \(\overline{XW}\) and \(\overline{WK}\) directions for a bare opal system. Labels refer to irreducible
representations of \(C_{2v}\) and \(C_{1h}\) respectively. Some inset figures
are included for the sake of clarity.

\textbf{Figure 5}. Detail of the photonic band structure normal to  the \{111\} set of
planes of a bare opal system. Notice that the maximum value of \(|k_{z}|\) is \(\sqrt{3}/2\)
in units of 2\(\pi\)/a. The thin horizontal lines define the frequencies at which the
branching ratios are obtained.
    
\begin{table}
\caption{The symmetry point groups at the corner of the irreducible
part of the Brillouin zone in the fcc lattice according to Schoenflies notation.
Notice that the components of representative vectors are in units of 2\(\pi\)/a}
{\centering \begin{tabular}{ccc}
Label&
Representative \textbf{k} vector&
Symmetry point group\\
\hline
\(\Gamma\)&
(0, 0, 0)&
\(O_{h}\)\\
\(X\)&
(1, 0, 0)&
\(D_{4h}\)\\
\(L\)&
(1/2, 1/2, 1/2)&
\(D_{3d}\)\\
\(U\)&
(1, 1/4, 1/4)&
\(C_{2v}\)\\
\(W\)&
(1, 1/2, 0)&
\(D_{2d}\)\\
\(K\)&
(3/4, 3/4, 0)&
\(C_{2v}\)\\
\end{tabular}\par}
\end{table}

\begin{table}
\caption{The irreducible representations corresponding to eigenvalues first to twentieth
at the corners of the irreducible part of the first Brillouin zone in a bare
opal system}

{\centering \begin{tabular}{|c|c|c|c|c|c|}
\( \Gamma [O_{h}] \)&
\( X[D_{4h}] \)&
\( L[D_{3d}] \)&
\( U[C_{2v}] \)&
\( W[D_{2d}] \)&
\( K[C_{2v}] \)\\
\hline 
T\( _{1u} \)&
E\( _{u} \)&
E\( _{g} \)&
B\( _{2} \)&
A\( _{1} \)&
B\( _{2} \)\\
T\( _{1g} \)&
E\( _{g} \)&
E\( _{u} \)&
B\( _{1} \)&
E&
B\( _{1} \)\\
E\( _{u} \)&
A\( _{2g} \)&
A\( _{1u} \)&
A\( _{1} \)&
B\( _{1} \)&
A\( _{1} \)\\
T\( _{2u} \)&
E\( _{g} \)&
E\( _{u} \)&
A\( _{2} \)&
A\( _{2} \)&
A\( _{2} \)\\
E\( _{g} \)&
B\( _{1g} \)&
E\( _{g} \)&
B\( _{1} \)&
E&
B\( _{1} \)\\
T\( _{1u} \)&
B\( _{1u} \)&
A\( _{1g} \)&
B\( _{2} \)&
B\( _{2} \)&
B\( _{2} \)\\
T\( _{2g} \)&
E\( _{u} \)&
E\( _{g} \)&
A\( _{2} \)&
A\( _{2} \)&
A\( _{2} \)\\
T\( _{2u} \)&
A\( _{2u} \)&
A\( _{2g} \)&
B\( _{2} \)&
B\( _{2} \)&
B\( _{2} \)\\
&
B\( _{2u} \)&
E\( _{u} \)&
A\( _{1} \)&
E&
A\( _{1} \)\\
&
A\( _{1u} \)&
A\( _{2u} \)&
B\( _{1} \)&
B\( _{1} \)&
B\( _{1} \)\\
&
B\( _{1g} \)&
A\( _{2g} \)&
B\( _{1} \)&
A\( _{1} \)&
B\( _{1} \)\\
&
E\( _{u} \)&
E\( _{g} \)&
B\( _{2} \)&
B\( _{1} \)&
B\( _{2} \)\\
&
E\( _{g} \)&
E\( _{u} \)&
A\( _{2} \)&
A\( _{1} \)&
A\( _{2} \)\\
&
B\( _{2g} \)&
&
A\( _{2} \)&
E&
A\( _{2} \)\\
&
&
&
B\( _{2} \)&
&
B\( _{2} \)\\
&
&
&
B\( _{1} \)&
&
B\( _{1} \)\\
&
&
&
A\( _{1} \)&
&
A\( _{1} \)\\
&
&
&
B\( _{2} \)&
&
B\( _{2} \)\\
&
&
&
A\( _{1} \)&
&
A\( _{1} \)\\
&
&
&
B\( _{1} \)&
&
B\( _{1} \)\\
\end{tabular} \par}
\end{table}  
\vspace{0.3cm}

{\begin{table}
 
\caption{The representations of incident plane waves and the subsequent uncoupled bands
along
some relevant high-symmetry directions in the Brillouin zone of the fcc lattice. Point
symmetry groups are described (in brackets) according to Schoenflies notation.}
 
{\centering \begin{tabular}{|c|c|c|}

Incidence Direction&
PW Representation&
Uncoupled Bands\\
\hline
\(\Gamma L [C_{3v}]\)&
E&
\(A_{1}, A_{2}\)\\
\(\Gamma X [C_{4v}]\)&
E&
\(A_{1}, A_{2}, B_{1}, B_{2}\)\\
\(\Gamma K [C_{2v}]\)&
\(B_{1}\oplus B_{2}\)&
\(A_{1}, A_{2}\)\\
\end{tabular}\par}
\end{table} \par}
   
\begin{table}
\caption{The compatibility relations of irreducible representations at \protect\( \Gamma \protect \) point
in the fcc lattice}

{\centering \begin{tabular}{|c|c|c|c|c|c|}
\( \Gamma [O_{h}] \)&
\( \overline{\Gamma X}[C_{4v}] \)&
\( \overline{\Gamma U}[C_{1h}] \)&
\( \overline{\Gamma L}[C_{3v}] \)&
\( \overline{\Gamma K}[C_{2v}] \)&
\( \overline{\Gamma W}[C_{1h}] \)\\
\hline 
A\( _{1g} \)&
A\( _{1} \)&
A'&
A\( _{1} \)&
A\( _{1} \)&
A'\\
\hline 
A\( _{2g} \)&
B\( _{1} \)&
A''&
A\( _{2} \)&
B\( _{2} \)&
A'\\
\hline 
E\( _{g} \)&
A\( _{1}\oplus  \)B\( _{1} \)&
A'\( \oplus  \)A''&
E&
A\( _{1}\oplus  \)B\( _{2} \)&
2A'\\
\hline 
T\( _{1g} \)&
E\( \oplus  \)A\( _{2} \)&
A'\( \oplus  \)2A''&
E\( \oplus  \)A\( _{2} \)&
A\( _{2}\oplus  \)B\( _{1}\oplus  \)B\( _{2} \)&
A'\( \oplus  \)2A''\\
\hline 
T\( _{2g} \)&
E\( \oplus  \)B\( _{2} \)&
2A'\( \oplus  \)A''&
E\( \oplus  \)A\( _{1} \)&
A\( _{1}\oplus  \)A\( _{2}\oplus  \)B\( _{1} \)&
A'\( \oplus  \)2A''\\
\hline 
A\( _{1u} \)&
A\( _{2} \)&
A''&
A\( _{2} \)&
A\( _{2} \)&
A''\\
\hline 
A\( _{2u} \)&
B\( _{2} \)&
A'&
A\( _{1} \)&
B\( _{1} \)&
A''\\
\hline 
E\( _{u} \)&
A\( _{2}\oplus  \)B\( _{2} \)&
A'\( \oplus  \)A''&
E&
A\( _{2}\oplus  \)B\( _{1} \)&
2A''\\
\hline 
T\( _{1u} \)&
E\( \oplus  \)A\( _{1} \)&
2A'\( \oplus  \)A''&
E\( \oplus  \)A\( _{1} \)&
A\( _{1}\oplus  \)B\( _{1}\oplus  \)B\( _{2} \)&
2A'\( \oplus  \)A''\\
\hline 
T\( _{2u} \)&
E\( \oplus  \)B\( _{1} \)&
A'\( \oplus  \)2A''&
E\( \oplus  \)A\( _{2} \)&
A\( _{1}\oplus  \)A\( _{2}\oplus  \)B\( _{2} \)&
2A'\( \oplus  \)A''\\
\end{tabular}\par}
\end{table} 

\vspace{0.3cm}
{\begin{table}

\caption{The compatibility relations of irreducible representations at \protect\( X\protect \) point
in the fcc lattice}

\begin{tabular}{|c|c|c|c|}
\( X[D_{4h}] \)&
\( \overline{\Gamma X}[C_{4v}] \)&
\( \overline{XU}[C_{2v}] \)&
\( \overline{XW}[C_{2v}] \)\\
\hline 
A\( _{1g} \)&
A\( _{1} \)&
A\( _{1} \)&
A\( _{1} \)\\
\hline 
A\( _{2g} \)&
A\( _{2} \)&
B\( _{2} \)&
B\( _{2} \)\\
\hline 
B\( _{1g} \)&
B\( _{1} \)&
B\( _{2} \)&
A\( _{1} \)\\
\hline 
B\( _{2g} \)&
B\( _{2} \)&
A\( _{1} \)&
B\( _{2} \)\\
\hline 
E\( _{g} \)&
E&
A\( _{2}\oplus  \)B\( _{1} \)&
A\( _{2}\oplus  \)B\( _{1} \)\\
\hline 
A\( _{1u} \)&
A\( _{2} \)&
A\( _{2} \)&
A\( _{2} \)\\
\hline 
A\( _{2u} \)&
A\( _{1} \)&
B\( _{1} \)&
B\( _{1} \)\\
\hline 
B\( _{1u} \)&
B\( _{2} \)&
B\( _{1} \)&
A\( _{2} \)\\
\hline 
B\( _{2u} \)&
B\( _{1} \)&
A\( _{2} \)&
B\( _{1} \)\\
\hline 
E\( _{u} \)&
E&
A\( _{1}\oplus  \)B\( _{2} \)&
A\( _{1}\oplus  \)B\( _{2} \)\\
\end{tabular}
\end{table} 
\vspace{0.3cm}

\vspace{0.3cm}
{\begin{table}

\caption{The compatibility relations of irreducible representations at \protect\( L\protect \) point
in the fcc lattice}

{\centering \begin{tabular}{|c|c|c|c|c|}
\( L[D_{3d}] \)&
\( \overline{\Gamma L}[C_{3v}] \)&
\( \overline{LW}[C_{2}] \)&
\( \overline{LU}[C_{1h}] \)&
\( \overline{LK}[C_{1h}] \)\\
\hline 
A\( _{1g} \)&
A\( _{1} \)&
A&
A'&
A'\\
\hline 
A\( _{2g} \)&
A\( _{2} \)&
B&
A''&
A''\\
\hline 
E\( _{g} \)&
E&
A\( \oplus  \)B&
A'\( \oplus  \)A''&
A'\( \oplus  \)A''\\
\hline 
A\( _{1u} \)&
A\( _{2} \)&
A&
A''&
A''\\
\hline 
A\( _{2u} \)&
A\( _{1} \)&
B&
A'&
A'\\
\hline 
E\( _{u} \)&
E&
A\( \oplus  \)B&
A'\( \oplus  \)A''&
A'\( \oplus  \)A''\\
\end{tabular}\par}
\end{table} 
\vspace{0.3cm}

\vspace{0.3cm}
{\begin{table}

\caption{The compatibility relations of irreducible representations at \protect\( U\protect \) point
in the fcc lattice}

{\centering \begin{tabular}{|c|c|c|c|c|}
\( U[C_{2v}] \)&
\( \overline{\Gamma U}[C_{1h}] \)&
\( \overline{XU}[C_{2v}] \)&
\( \overline{LU}[C_{1h}] \)&
\( \overline{UW}[C_{1h}] \)\\
\hline 
A\( _{1} \)&
A'&
A\( _{1} \)&
A'&
A'\\
\hline 
A\( _{2} \)&
A''&
A\( _{2} \)&
A''&
A''\\
\hline 
B\( _{1} \)&
A'&
B\( _{1} \)&
A'&
A''\\
\hline 
B\( _{2} \)&
A''&
B\( _{2} \)&
A''&
A'\\
\end{tabular}\par}
\end{table} \par}
\vspace{0.3cm}

\vspace{0.3cm}
{\begin{table}

\caption{The compatibility relations of irreducible representations at \protect\( W\protect \) point
in the fcc lattice}

{\centering \begin{tabular}{|c|c|c|c|c|c|}
\( W[D_{2d}] \)&
\( \overline{\Gamma W}[C_{1h}] \)&
\( \overline{XW}[C_{2v}] \)&
\( \overline{LW}[C_{2}] \)&
\( \overline{UW}[C_{1h}] \)&
\( \overline{WK}[C_{1h}] \)\\
\hline 
A\( _{1} \)&
A'&
A\( _{1} \)&
A&
A'&
A'\\
\hline 
A\( _{2} \)&
A''&
A\( _{2} \)&
B&
A''&
A''\\
\hline 
B\( _{1} \)&
A''&
A\( _{2} \)&
A&
A''&
A''\\
\hline 
B\( _{2} \)&
A'&
A\( _{1} \)&
B&
A'&
A'\\
\hline 
E&
A'\( \oplus  \)A''&
B\( _{1}\oplus  \)B\( _{2} \)&
A\( \oplus  \)B&
A'\( \oplus  \)A''&
A'\( \oplus  \)A''\\
\end{tabular}\par}
\end{table} \par}
\vspace{0.3cm}
\vspace{0.3cm}
{\begin{table} 

\caption{The compatibility relations of irreducible representations at \protect\( K\protect \) point
in the fcc lattice}

{\centering \begin{tabular}{|c|c|c|c|}
\( K[C_{2v}] \)&
\( \overline{\Gamma K}[C_{2v}] \)&
\( \overline{LK}[C_{1h}] \)&
\( \overline{WK}[C_{1h}] \)\\
\hline 
A\( _{1} \)&
A\( _{1} \)&
A'&
A'\\
\hline 
A\( _{2} \)&
A\( _{2} \)&
A''&
A''\\
\hline 
B\( _{1} \)&
B\( _{1} \)&
A'&
A''\\
\hline 
B\( _{2} \)&
B\( _{2} \)&
A''&
A'\\
\end{tabular}\par}
\end{table} \par}

\end{document}